\documentclass[letterpaper, 10 pt, conference]{ieeeconf}  
\usepackage[english]{babel}
\usepackage{amsmath}
\usepackage{graphicx}
\graphicspath{{figures/}}
\usepackage[colorlinks=true, allcolors=blue]{hyperref}
\usepackage{bm}
\usepackage{upgreek}
\usepackage{subcaption}
\usepackage{caption}
\usepackage[font=small]{caption}
\usepackage{soul}

\IEEEoverridecommandlockouts                              

\title{\LARGE \bf
Behavior-Aware Online Prediction of Obstacle Occupancy using Zonotopes}

\author{Alvaro Carrizosa-Rendon$^{1,2}$, Jian Zhou$^{3}$, Erik Frisk$^{4}$, Vicenç Puig$^{1,2}$ and Fatiha Nejjari$^{2}$ 
\thanks{*This work has been co-financed by the Spanish State Research Agency (AEI) and the European Regional Development Fund (ERFD) through the project SaCoAV (ref. MINECO PID2020-114244RB-I00)  and by the DGR of Generalitat de Catalunya (SAC group ref. 2021/SGR/278). J. Zhou and E. Frisk acknowledge the support by Strategic Research Area at Link\"oping-
Lund in Information Technology (ELLIIT).}
\thanks{$^{1}$Institut de Rob\`otica i Inform\`atica Industrial CSIC-UPC, Carrer Llorens Artigas, 4-6, Barcelona, 08028, Spain.}%
\thanks{$^{2}$Research Center for Supervision, Safety and Automatic Control (CS2AC), Universitat
Polit\`ecnica de Catalunya (UPC), Rambla Sant Nebridi 22, 08222 Terrassa, Spain.}%
\thanks{$^{3}$AVATR Co., Ltd., Choingqing, China. Email: jian.zhou3@avatr.com}
\thanks{$^{4}$Department of Electrical, Engineering Link\"oping University, Sweden. Email: erik.frisk@liu.se}%
}

\begin{document}

\maketitle
\thispagestyle{empty}
\pagestyle{empty}

\begin{abstract} 
Predicting the motion of surrounding vehicles is key to safe autonomous driving, especially in unstructured environments without prior information. This paper proposes a novel online method to accurately predict the occupancy sets of surrounding vehicles based solely on motion observations. The approach is divided into two stages: first, an Extended Kalman Filter and a Linear Programming (LP) problem are used to estimate a compact zonotopic set of control actions; then, a reachability analysis propagates this set to predict future occupancy.
The effectiveness of the method has been validated through simulations in an urban environment, showing accurate and compact predictions without relying on prior assumptions or prior training data.
\end{abstract}

\section{INTRODUCTION}
Autonomous driving has generated great research interests given the expected benefits, such as reducing accidents, optimizing traffic efficiency and energy management~\cite{yuchen2024impactCAV}. However, ensuring safety remains a major challenge, particularly in urban environments, where multiple agents interact dynamically~\cite{hacohen2022surveyAD}.
Predicting the motion of surrounding vehicles (SVs) is critical to designing safe motion planning and control strategies for autonomous vehicles. Motion-prediction methods can be broadly categorized into probabilistic or set-based approaches, which are further implemented through model-based, learning-based, or integrated strategies~\cite{liu2023radius}. Despite their proven effectiveness in various scenarios, a significant research problem remains in predicting the motion of dynamic SVs without prior information such as training datasets reflecting the driving behavior or road-geometry details. This becomes critical in urban scenarios like intersections, where the ego-vehicle (EV) has limited information and short interaction time. An efficient online prediction method is therefore required.

This paper addresses the challenge of online prediction of future occupancy sets of SVs by proposing a set-based method that requires no prior information about the SVs or their environment. For any SV under observation, the method uses online estimations of its control behavior to construct a Control-Input set. This set is then used to compute future occupancy through a zonotopic reachability analysis, exploiting the compactness and efficiency of zonotopes. The main contributions of this work are:
\begin{enumerate}
 \item A novel approach to predict the SV’s occupancy without prior knowledge or assumptions, by estimating its behavior online and applying a reachability analysis based on the expected control actions.
 \item A computationally efficient approach to estimate the SV's Control-Input set, combining a nonlinear observer (EKF) and a linear programming (LP) problem to define an optimized zonotope that captures the observed information of the SV in real time.
\item An alternative efficient formulation for performing a forward reachability analysis based on the zonotopic Control-Input set, posed as the analogous problem of propagating uncertainties around a nominal trajectory.
\end{enumerate}


\section{RELATED WORKS} \label{sec:related_work}
Given a predefined motion model, a key challenge in predicting the motion of obstacles lies in capturing their uncertain control actions. Since these are typically unknown to the EV, a common strategy, often referred to as the formal approach, is to propagate the reachable set under worst-case assumptions. Such assumptions have been used in \cite{batkovic2023experimental} to predict pedestrian occupancy for a robust motion planner. 
The work in \cite{koschi2021prediction} proposed a formal set-based prediction method for various objects, considering the influence of traffic rules and hidden traffic participants (occlusions). Although formal approaches guarantee safety by covering all possible maneuvers, they tend to be overly conservative, limiting planning feasibility or performance.

To mitigate the conservatism of formal methods, several approaches have been proposed. For example, \cite{gao2021risk} reduces the size of the reachable set in overtaking scenarios using the supermartingale assumption. The work in~\cite{zhou2024robust} considered both the uncertain longitudinal and lateral control actions of the obstacle, which was described by a linear model, and the occupancy was propagated based on H-polytope. This idea was further extended in~\cite{zhou2025robust}, where the road-geometry information was integrated into the reachability predictions in cornering scenarios like the roundabout. The method proposed in this paper differs from these relevant works in three main aspects: (1) in contrast to~\cite{gao2021risk}-\cite{zhou2025robust}, the approach presented in this work models SVs nonlinearly to better capture their dynamics; (2) compared to \cite{gao2021risk} and \cite{zhou2024robust}, the proposed approach achieves higher computational efficiency via zonotopes; and (3) the proposed method does not require additional information such as the road geometry as in~\cite{zhou2025robust} for reachability prediction in cornering scenarios.

In addition to formal methods and approaches to reduce conservatism, another relevant research direction is integrating learning-based strategies into estimation filters. For example, the approach presented in~\cite{xu2021trafficEKF} combines an Extended Kalman Filter with a deep learning model to estimate control actions of SVs based on environmental perception. In contrast, the present work applies an EKF based solely on online motion observations, without relying on learned models or offline training. This characteristic enhances its applicability to unknown or unstructured environments.

To improve the computational efficiency of reachability prediction, there is a growing tendency toward propagating zonotopes, a particular type of polytope. As described in~\cite{althoff2021set}, zonotopes enable efficient propagation over a time horizon as well as a compact mathematical representation that reduces the complexity of set operations. Applications of this approach can be found in~\cite{liu2024refine} 
which used the reachable set based on zonotopes to study the future evolutions of the system accurately. In these approaches, zonotopes were used for safe motion planning. Given the desired accuracy, the computation of possible reachable sets was performed offline, but then combined with online optimization techniques to ensure obstacle avoidance. 

\section{PROBLEM STATEMENT} \label{sec:problem_statement}
This work addresses the problem of predicting the future occupancy sets of SVs in urban environments by continuously updating both the behavior estimation and the prediction of the occupancy sets based solely on recent motion observations. The method is designed to operate in both structured and unstructured scenarios, assuming that SVs follow typical urban-like driving behaviors.The method is explicitly designed to operate without any prior information about future control actions or maneuver intentions, relying solely on position and velocity measurements.\\
To achieve accurate and adaptable predictions under these conditions, the method solves two sub-problems iteratively:
\begin{enumerate}
    \item \textbf{Behavior estimation of the SV:} Without prior knowledge of future control actions, the first challenge is to infer the SV’s behavior from noisy and uncertain measurements of its observable states. From these observations, a set representing the observed behavior has to be efficiently defined.
    \item \textbf{Prediction of the occupancy sets:} Based on the estimated behavior, the second challenge is to predict the SV’s future states over a fixed horizon. For this, a Reachability Analysis is proposed to compute all possible evolutions and extract the corresponding occupancy sets at each step.
\end{enumerate}
The method combines optimal observers, optimization techniques, and set-based approaches. Zonotopes have been specifically selected due to their compactness and computational efficiency, enabling continuous updates and accurate predictions while maintaining a non-conservative approach. 

\section{METHODOLOGY} \label{sec:methodology}
\begin{figure}[!htb]
\centering
\includegraphics[width=\columnwidth]{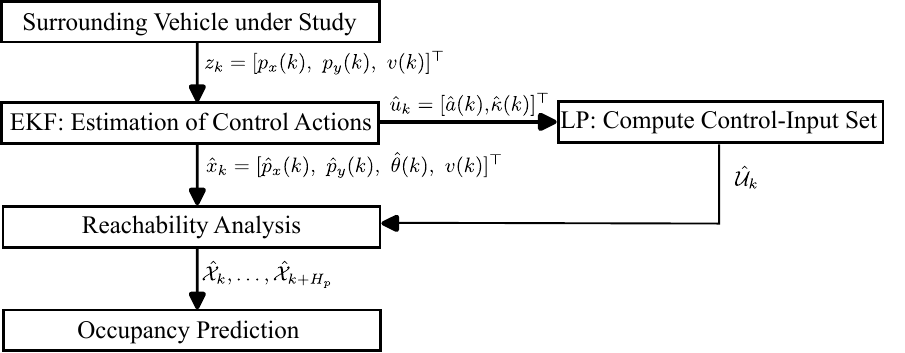}
\caption{Methodology stages: (1) estimate the Control-Input set $\mathcal{\hat{U}}_k$ of the SV from observed inputs $\hat{u}_k$ via EKF+LP; (2) predict the occupancies $\mathcal{\hat{X}}_k, \ldots, \mathcal{\hat{X}}_{k+N_p}$ by zonotopic reachability.}
\label{fig:scheme}
\end{figure}

The methodology addresses the two sub-problems from Section~\ref{sec:problem_statement} through two sequential and iterative stages: estimation of the SV’s current behavior from recent observations, and prediction of its future states to extract occupancy sets. These stages are repeated as new measurements become available. The steps of each stage are illustrated in Figure~\ref{fig:scheme}.\\
As shown in the figure, the first stage includes two steps: an Extended Kalman Filter (EKF) based on a nonlinear vehicle model to estimate the control actions, and a Linear Programming (LP) problem to compute the Control-Input set, represented as a zonotope. In the second stage, a zonotopic reachability analysis is proposed to predict the SV’s states over the horizon, from which occupancy sets are extracted and simplified to eliminate redundancies.\\

It is worth noting that the methodology combines optimal observers and optimization techniques to compute the Control-Input set, enabling accurate estimations based on recent observations of the SV. Moreover, the set is parameterized as a zonotope, providing a compact mathematical representation and allowing efficient computation, as discussed in~\cite{althoff2021set}, which also facilitates the forward reachability propagation. 
\subsection{Model of the Surrounding Vehicles}
The first stage in Figure~\ref{fig:scheme} aims to define a set representing the estimated behavior of the SVs based on observations. In this paper, the estimations are based on the single-track kinematic model \cite{rajamani2011vehicle} described with equations~\eqref{eq:cinematic_model1}--\eqref{eq:cinematic_model4}.
\begin{align}\label{eq:cinematic_model1}
    \dot{p}_x(t) &= v(t)\cdot \cos(\theta(t))\\
    \dot{p}_y(t) &= v(t)\cdot \sin(\theta(t))\\
    \dot{\theta}(t) &= v(t) \cdot \frac{\tan(\delta(t))}{L} = v(t)\cdot \kappa(t) \label{eq:cinematic_model3}\\
    \dot{v}(t) &= a(t) \label{eq:cinematic_model4}
\end{align}

The states of the model are the longitudinal and lateral positions and orientation of the vehicle $\left(p_x(t),p_y(t),\theta(t)\right)$ and its linear velocity $\left(v(t)\right)$. The control inputs are both acceleration $\left(a(t)\right)$ and steering angle $\left(\delta(t)\right)$. As the wheelbase length  $\left(L\right)$ is unknown, it is combined with the steering angle in equation~(\ref{eq:cinematic_model3}) as the curvature of the trajectory performed $\left(\kappa(t)\right)$. Consequently, the control actions to be estimated are $a(t)$ and $\kappa(t)$.
\subsection{Estimation of the Control Actions}
To estimate the control actions in model~\eqref{eq:cinematic_model1}--\eqref{eq:cinematic_model4}, an EKF is applied with an approach formulated similar to the one presented in~\cite{xu2021trafficEKF}, but adapted to different measurable states and with a distinct objective. For implementation, the model~\eqref{eq:cinematic_model1}--\eqref{eq:cinematic_model4} is reformulated as an augmented system, introducing the acceleration and curvature as additional states. The system is then discretized by Euler forward approach:
\begin{align} 
    p_{x}(k+1) &= p_x(k) + v(k)\cdot \cos(\theta(k)) \cdot T_s \label{eq:augmented_disc1}\\
    p_{y}(k+1) &= p_y(k) + v(k)\cdot \sin(\theta(k)) \cdot T_s\\
    {\theta}(k+1) &= \theta(k)+ v(k)\cdot \kappa(k)\cdot T_s\\
    {v}(k+1) &= v(k) + a(k)\cdot T_s \label{eq:augnemtned_disc4}\\ 
    {a}(k+1) &= a(k) + \omega_1(k)\\
    \kappa(k+1) &= \kappa(k) +\omega_{2}(k) \label{eq:augmented_disc6}
\end{align}

\noindent $T_s$ denotes the sampling interval, $k$ is the discrete time index. The terms $\omega_1(k)$ and $\omega_2(k)$ are explicitly introduced as disturbances to model the variations in the SV's control actions caused by changes in its behavior due to dynamic traffic conditions.
For the augmented system described in~\eqref{eq:augmented_disc1}--\eqref{eq:augmented_disc6}, the measured information is the longitudinal and lateral positions and the velocity of the SV at the current time step $k$. These are collected in the measurement vector $z(k) = [p_x(k) \ p_y(k) \ v(k)]^{\top}$, which is affected by measurement noise. Based on this information and the structure of the model, the EKF is applied to estimate ${a}(k)$ and $\kappa(k)$ at every time step $k$. These estimations are then used to construct the Control-Input set in the next stage.
\subsection{Estimation of the Control-Input Set of SV}\label{sec: set estimation}
Based on the control actions estimated by the EKF at each time step, the set of accelerations and curvatures can be computed using a parameterized approach. In this work, the control input set is parameterized as an optimal zonotope constructed from the recently observed control actions of the SV. Following this idea, the computation is structured in four main steps:
\begin{enumerate}
\item \textbf{Select the last $N$ observed control actions with a sliding window}:Limiting the horizon to the most recent control actions reduces computational cost and avoids including regions from former maneuvers that no longer represent the current behavior.
\item \textbf{Define a primitive generator matrix with unitary generators}: The used generator array consists of angularly equispaced vectors covering all orientations. 
\item \textbf{Compute the Control-Input set based on the $N$ control actions}: This is achieved by solving a linear programming (LP) problem to determine the optimal zonotope containing all selected observations. 
\item \textbf{Expansion of the zonotope to increase accuracy}: To ensure that the Control-Input set accounts for small variations in the observed control actions, two additional small generators, $g_{u_1}$ and $g_{u_2}$, are added in the direction of each control action. This acts as a safety margin, increasing conservatism but ensuring that slight changes remain within the set. The margin should be based on the typical covariance of the observations during stationary behavior, with a small extra buffer to enhance safety. However, it must be carefully chosen to avoid excessive conservatism, which may reduce reliability in dynamic scenarios.
\end{enumerate}
The aim of the LP problem is to find the optimal center and shape of the zonotope that represents the estimated control input set of the SV. The LP problem is formulated as in~\cite{bravo2004control}: 
\begin{align}\label{eq:opt_zonotope}
    &\min_{c_u,\alpha_i,\delta_{i,j}} \hspace{2mm} \sum_{i=1}^{n_g} \alpha_i\\
    &\text{s.t. \hspace{4mm}} \small{u_j = c_u + \sum_{i=1}^{n_g} \delta_{i,j} \cdot g_i \label{eq:cons_control}}\\
     &\hspace{9mm}  \small{\alpha_i \geq 0  \label{eq:pos_alpha} \;\;\forall i}\\
      &\hspace{5mm}\small{-\alpha_i \leq \delta_{i,j} \leq \alpha_i \;\;\forall i,j}\label{eq:cons_alpha}
\end{align}
\noindent Where, $c_u$ is the center of the zonotope and $\alpha_i$ the scaling factors for the $n_g$ predefined unitary generators $g_i$, both to be optimized. The auxiliary variables $\delta_{i,j}$ ensure that all the selected observations $u_j$ are enclosed within the resulting zonotope through constraints (\ref{eq:cons_control})-(\ref{eq:cons_alpha}).
To keep the formulation simple and computationally efficient, the chosen cost function~(\ref{eq:opt_zonotope}) minimizes the sum of the scaling factors $\alpha_i$, which corresponds to minimizing the spread of the zonotope, leading to a compact representation of the control input set.
By this way, the set of control input set $\mathcal{\hat{U}}$ is defined by the optimized center $c_u$ and the generator matrix $H_u$. As shown in~(\ref{eq:generator}), it is composed of the $n_g$ predefined generators $g_i$ each scaled by its optimized factor $\alpha_i$, along with two additional generators $g_{u_1}$ and $g_{u_2}$, appended as extra columns to expand the zonotope and enhance the robustness of the method: 
\begin{equation} \label{eq:generator} 
    H_u = [\alpha_1 \cdot g_1, ..., \alpha_{n_g} \cdot g_{n_g} | g_{u_1} | g_{u_2}]
\end{equation}

\subsection{Reachability Analysis}
After estimating the Control-Input set $\mathcal{\hat{U}}$, the next step consists in predicting the forward occupancy sets of the SV based on it. This is achieved by performing a forward reachability analysis~\cite{althoff2021set}. For that, all possible values that the states~(\ref{eq:augmented_disc1})--(\ref{eq:augnemtned_disc4}) can take at each time instant are computed, taking as initial values the estimations of the EKF, and as a set of applicable control actions, the most recent estimations of the Control-Input set $\mathcal{\hat{U}}_k$. 

Based on the properties of the zonotopes, this procedure can be formulated as an iterative step-by-step process, decoupling the propagation of the center and the computation of the new generator with the expressions~\eqref{eq:propag_c} and \eqref{eq:propag_H}, which formulates it as the analog problem of computing a nominal path and propagating uncertainties of the control action around it. 
Equation~\eqref{eq:propag_c} makes reference to the computation of the nominal trajectory along the prediction horizon obtained from equations~(\ref{eq:augmented_disc1})-(\ref{eq:augnemtned_disc4}) particularized with the center of the predicted zonotope in the previous steps and the center of the Control-Input set. As prediction of the first step, the observations of the EKF are taken. 

A common way to propagate the generator matrix around the center of the zonotope is by computing the linearized system and input matrices around the operating point. Since it is not desirable to work just around a single operating point missing feasible evolutions of the SV, the system and input matrices ($\mathcal{A}_k$ and $\mathcal{B}_k$) have been formulated as interval matrices to capture all the possible evolutions of the system based on linearizations over all the states contained in the set of possible states.
\begin{align}
c_x(k+1) &= f\left(c_x(k),c_u \right) \label{eq:propag_c}\\
\mathcal{H}_x(k+1) &= \mathcal{A}_k \cdot H_x(k) + \mathcal{B}_k \cdot H_u \label{eq:propag_H}
\end{align}
In other words, the interval matrices used for the propagation of the set are matrices whose terms are intervals computed according to expressions~$\bm{\rho_{{ij}_k}}$ in order to take into account all the possible matrices which could be obtained from linearizing around every possible operating point. The interval terms $\bm{\rho_{{ij}_k}}$ are computed considering all possible values of the states and control inputs according to interval arithmetic where states $\bm{\uptheta}(\mathbf{k})$, $\mathbf{v(k)}$ and input $\bm{\upkappa}{(\mathbf{k})}$ are also intervals.
\begin{align}
    \mathbf{\mathcal{A}_k} &= \begin{bmatrix}
        1 & 0 & [{\rho}_{13_k}^{\rm min},{\rho}_{13_k}^{\rm max}] & [{\rho}_{14_k}^{\rm min},{\rho}_{14_k}^{\rm max}] \\
        0 & 1 & [{\rho}_{23_k}^{\rm min},{\rho}_{23_k}^{\rm max}] & [{\rho}_{24_k}^{\rm min},{\rho}_{24_k}^{\rm max}]\\
        0 & 0 & 1 & [{\rho}_{34_k}^{\rm min},{\rho}_{34_k}^{\rm max}]\\
        0 & 0 & 0 & 1
    \end{bmatrix} \label{eq:interval_A}\\
    \mathcal{B}_k &= \begin{bmatrix}
        0 & 0 & [\rho_{b_k}^{\rm min},\rho_{b_k}^{\rm max}] & 0  \\
        0 & 0  & 0 & T_s
    \end{bmatrix}^\top \label{eq:interval_B}
\end{align}
\vspace{-4mm}
\begin{align*}
    &\small{\bm{\rho_{13_k}} =-T_s \cdot \mathbf{v(k)} \cdot \sin(\bm{\uptheta}(\mathbf{k}))}
    &\small{\bm{\rho_{b_k}} = T_s \cdot \mathbf{v(k)}} \\
    &\small{\bm{\rho_{23_k}} = T_s \cdot \mathbf{v(k)} \cdot \cos(\bm{\uptheta}(\mathbf{k}))}  &\small{\bm{\rho_{34_k}} = T_s \cdot \bm{\upkappa}{(\mathbf{k})}} \\
    &\small{\bm{\rho_{14_k}} = T_s \cdot \cos(\bm{\uptheta}(\mathbf{k})) }
    &\small{\bm{\rho_{24_k}} = T_s \cdot \sin(\bm{\uptheta}(\mathbf{k}))}
\end{align*}

In the same way, the computed reachable set is not a single zonotope but rather a family of zonotopes due to the fact of being defined by a family of generator matrices as a consequence of the use of interval matrices presented in~(\ref{eq:propag_H}). Thus, an inclusion operation denoted by the symbol~$\diamond$ is performed as shown in equation~(\ref{eq:propag_H2}). The aim of this mathematical operation is to define a generator matrix that encloses the whole family. More information about the inclusion process, which consists of computing a single zonotope containing a whole family of zonotopes defined by an interval matrix of generators, can be consulted in \cite{alamo2005guaranteed}. As a result of this inclusion procedure, a resulting set defined by $c_x(k+1)$ and $H_x(k+1)$ is obtained, which contains the exact reachable set at time instant $k$ using the previously computed Control-Input set as  all possible control actions.
\begin{align}
    H_x(k+1) &= \diamond \mathcal{H}_x(k+1) \supseteq \mathcal{H}_x(k+1) \label{eq:propag_H2}
\end{align}

\subsection{Occupancy Set Extraction}
The final step for determining the occupancy sets $\mathcal{\hat{X}}_k$ consists of extracting them from the zonotopes computed during the reachability analysis. These zonotopes are defined in four dimensions, corresponding each of them to one state. Therefore, extracting solely the two first dimensions, the expected Cartesian locations of the SV is obtained. Since the projected zonotope is now two-dimensional, the dimensionality of its latent space can be reduced. Specifically, the projection is obtained by selecting the first and second components of each original generator. This may result in redundant generators, which can be debugged, either removed or merged, without altering the shape of the set. The simplification is carried out through the following steps:
\begin{enumerate}
    \item \textbf{Deletion of null generators:} It might be the case that many of the generators are null because their terms were referring to the two other remaining states,  therefore, the null generators can be removed without affecting the original shape.
    \item \textbf{Merge parallel generators:} Generators with the same direction can be merged into a single one, with a norm equal to the sum of their original norms. This simplifies the representation of the set while preserving its shape.
    \item \textbf{Reduction of the number of generators:} If the number of generators exceeds a predefined threshold, a more efficient zonotope can be computed by replacing the least influential generators with a bounding box that over-approximates their contribution. 
    \item \textbf{Dilation of the computed zonotope:} To increase safety, the occupancy set should be partially dilated to ensure that short-term predictions contain the SV's location, even if its behavior changes. The dilation magnitude should be tuned based on the system dynamics, balancing prediction accuracy and conservatism. A larger dilation extends the number of time steps with containment guarantees but increases conservatism; a smaller dilation yields tighter sets but requires more frequent updates to remain reliable. 
\end{enumerate}
At this stage, the predicted occupancy sets ($\mathcal{X}_{k,...,N_p}$) have been computed over the entire prediction horizon, balancing precision and conservatism. These sets delimit the area in which the observed vehicle is expected to be located at each future time step. 
\section{SIMULATION RESULTS} \label{sec:results}
\begin{figure}[hbt!]
\centering
\includegraphics[width=\columnwidth]{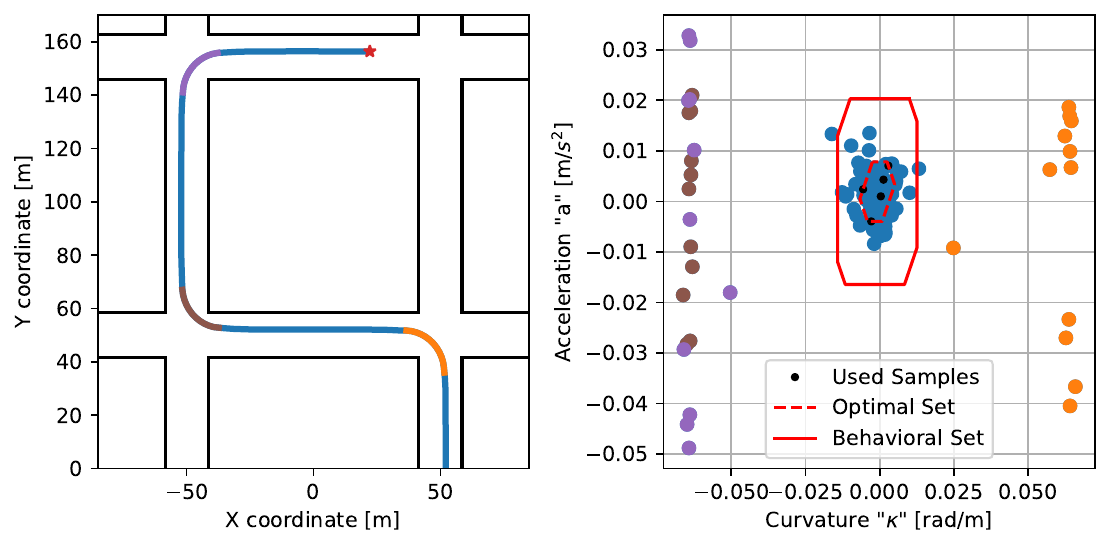}
\caption{\textbf{Left}: SV trajectory during $30 {\, s}$. \textbf{Right}: Observed control actions; black: samples used to fit the last Control-Input set; dashed red: optimized zonotope; solid red: dilated Control-Input set.}
\label{fig:path_followed}
\end{figure}

The proposed methodology was evaluated in simulations. The SV used a nonlinear single-track kinematic model with an MPC-based trajectory-tracking controller, implemented as in~\cite{zhou2024robust}. Gaussian noise was added to actions and measurements to simulate tracking and observation errors.  
The simulation scenario, shown in the left panel of Figure~\ref{fig:path_followed}, is a grid where the vehicle moves freely, executing urban-like maneuvers. The resulting path followed by the SV is represented with the multicolored curve in the image, ending at the point marked with a red star. The color coding differentiates between straight segments and turns performed by the vehicle. 
 The presented results corresponds to executing the method 150 iterations with a sampling interval of $T_s=0.2 \ {\rm s}$, a sliding window of $5$ samples, and $3$ generators to establish the Control-Input zonotope. The prediction horizon $N_p$ was set at $10$ samples for the study. The implementation was in Python, using package $uncertainties$ \cite{lecroart2024uncertainties} to manage them in propagation, CasADi \cite{Andersson2019CasADi} for formulating the LP problem and $qpOASES$ \cite{Ferreau2014} for optimizing the Control-Input set. 
\subsection{Behavior Estimation}
\begin{figure}[hbt!]
\centering
\includegraphics[width=\columnwidth]{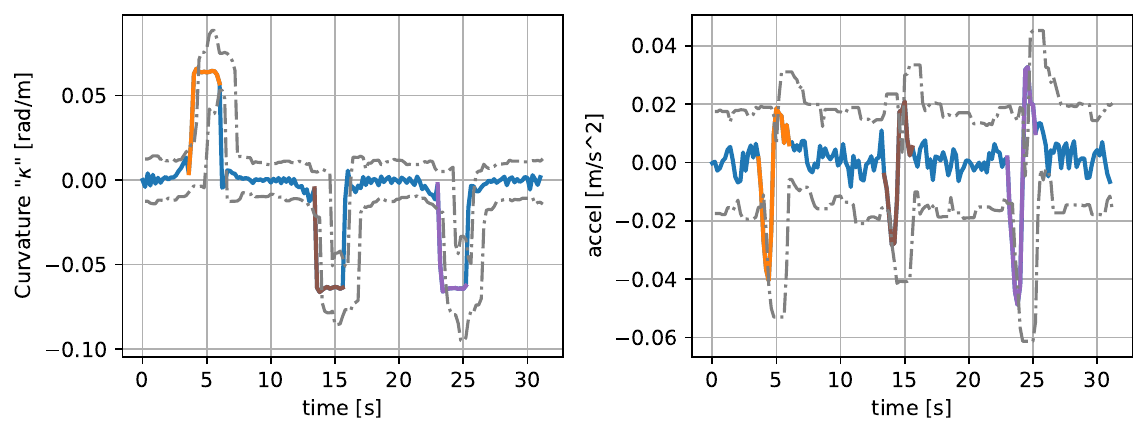}
\caption{Observed control actions during the simulation(colored by path segments as in  Figure~\ref{fig:path_followed}) and estimated Control-Input set bounds (gray dashed) at each time $k$.}
\label{fig:control_evolution}
\end{figure}

As demonstrated in Figure~\ref{fig:path_followed}, a small set of recent samples, combined with a proper dilation parameter, is able to represent the SV's behaviors. This is further illustrated in Figure~\ref{fig:control_evolution}, where, the observed control actions are color-coded according to the segment of the path. Additionally, the maximum and minimum values of the estimated Control-Input set of the SV are shown in dashed gray. These results clearly show how the zonotope effectively adjusts and accurately represents the vehicle's behavior. When the maneuver continues over time, the size of the zonotope decreases, precisely defining the observed actions. However, if a sudden behavioral change occurs, the observation falls outside the set, causing an expansion and adaptation of the shape. As consistent observations follow, it contracts again, refining the shape according to the new behavior.
\begin{figure}[hbt!]
\centering
\includegraphics[width=0.85\columnwidth]{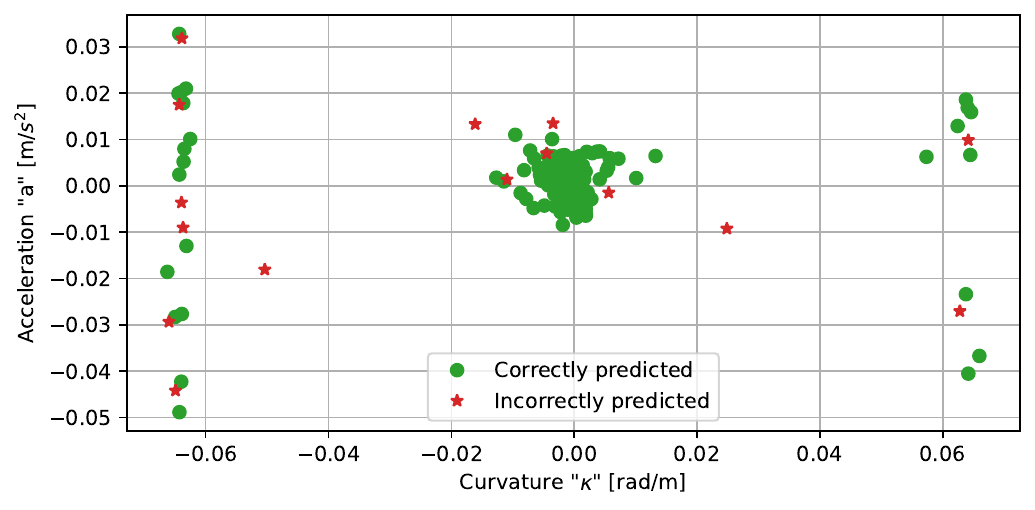}
\caption{Classification of observed control actions over $150$ iterations: green = inside the estimated Control-Input set; red = outside.}
\label{fig:control_pred}
\end{figure}
By analyzing the graph and comparing it with the details in Figure~\ref{fig:control_evolution}, it becomes evident that the magnitudes which were inaccurately predicted predominantly correspond to sudden behavioral changes and specific observations that lie at the extremes of the distribution. However, given the fast adaptation of the method in subsequent iterations, this will not be a critical aspect, as will be detailed in the following subsections.
In addition, to verify the correct functioning of the behavior estimation stage, an analysis was conducted at each iteration to ascertain whether the observed control action was within the estimated control input set. Over 150 iterations (Figure~\ref{fig:control_pred}), a success rate of $90.00\%$ was obtained. 

\subsection{Occupancy Prediction}
The accuracy of the predicted occupancy sets was evaluated comparing each prediction to the actual planned position of the SV at each time step. Throughout the simulation, $89.13\%$ of the $1500$ occupancy sets  were correct. 
Errors occur mainly at maneuver transitions, being more frequent at longer horizons, as the change in behavior has not yet been captured. This is not critical, as the method is executed iteratively, allowing predictions to be updated before reaching the time steps where they might no longer be valid. Table~\ref{tab:success} summarizes success versus horizon.
Additionally, average per-iteration time (EKF+LP+Reachability) is reported in Table~\ref{tab:computational}. Results show how cost increases with length of the prediction horizon, but remaining computationally efficient as a result of leveraging the properties of zonotope. \vspace{1mm}
\begin{table}[!ht]
\caption{Successful Rate Predicting the Occupancy Sets}
\label{tab:success}
\small
\begin{center}
\begin{tabular}{|c|c|c|c|c|c|}
\hline
\textbf{$N_p$} & 1 & 2 & 3 & 4 & 5 \\
\hline
\textbf{Success} &100\%&100\%&96.67\%&93.33\% & 90.67\%\\
\hline
\hline
 \textbf{$N_p$}&6 & 7 & 8 & 9 & 10\\
 \hline
 \textbf{Success}  &87.33\%& 85.33\%&81.33\% & 80.00\% & 76.67\%\\
\hline
\end{tabular}\\
\vspace{2mm} 
    \parbox{1\linewidth}{\scriptsize *Prediction success rate (\%) versus prediction step $N_p$.}
\end{center}
\end{table}\vspace{-1mm}
\begin{table}[!ht]
\caption{Computational Cost vs. Prediction Horizon}
\label{tab:computational}
\small
\begin{center}
\begin{tabular}{|c|c|c|c|c|}
\hline
\textbf{$N_p$} & 3 & 4 & 5 & 6 \\
\hline
\textbf{Com. Time [${\rm s}$]} &0.00295&0.00467&0.00727&0.00987\\
\hline
\hline
 \textbf{$N_p$}& 7 & 8 & 9 & 10\\
 \hline
 \textbf{Com. Time [${\rm s}$]} &0.01339&0.01751& 0.02224&0.02714\\
\hline
\end{tabular}\\
\vspace{2mm} %
    \parbox{1\linewidth}{\scriptsize *Average per-iteration computation time (s) versus prediction horizon $N_p$. 150 iterations; behavioral set of 3 generators and a 3-sample window.}
\end{center}
\end{table}
\subsection{Graphical Comparison with a Conservative Strategy}
\begin{figure}[ht!]
    \centering
    \begin{subfigure}[t]{0.31\columnwidth}
        \centering
        \includegraphics[width=\linewidth]{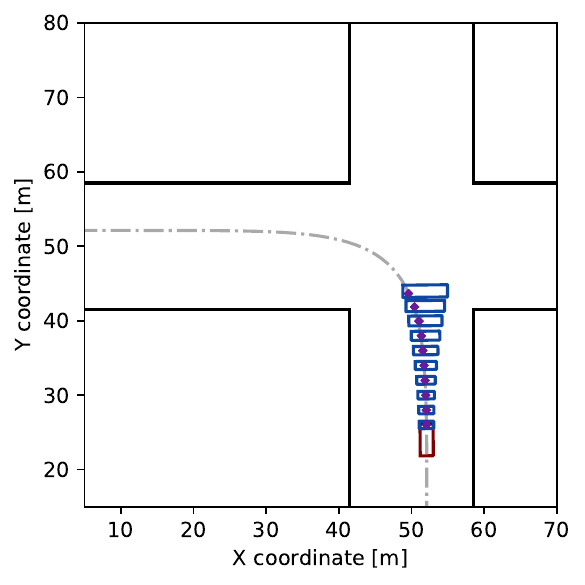}
        \caption{k = 12}
        \label{fig:scenario_12}
    \end{subfigure}
    \hfill
    \begin{subfigure}[t]{0.31\columnwidth}
        \centering
        \includegraphics[width=\linewidth]{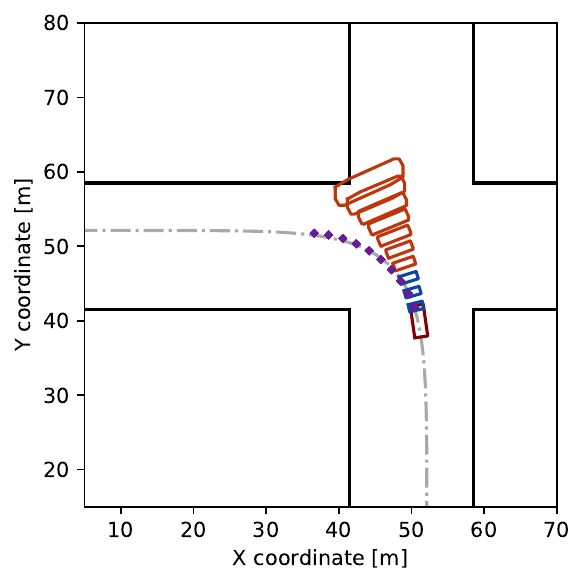}
        \caption{k = 20}
        \label{fig:scenario_20}
    \end{subfigure}
    \hfill
    \begin{subfigure}[t]{0.31\columnwidth}
        \centering
        \includegraphics[width=\linewidth]{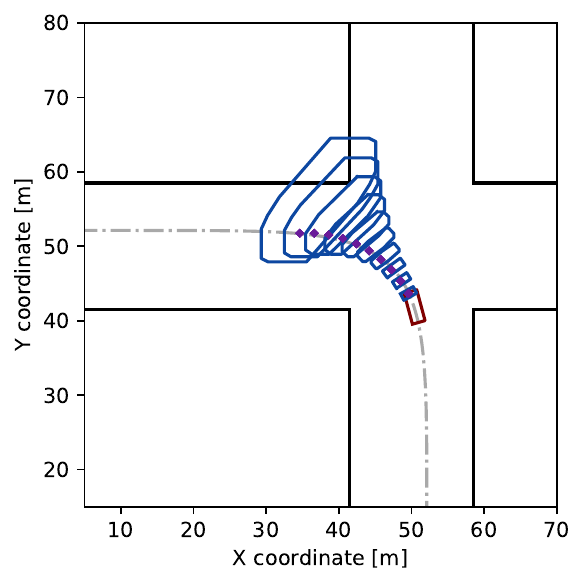}
        \caption{k = 21}
        \label{fig:scenario_21}
    \end{subfigure}
    \begin{subfigure}[t]{0.31\columnwidth}
        \centering
        \includegraphics[width=\linewidth]{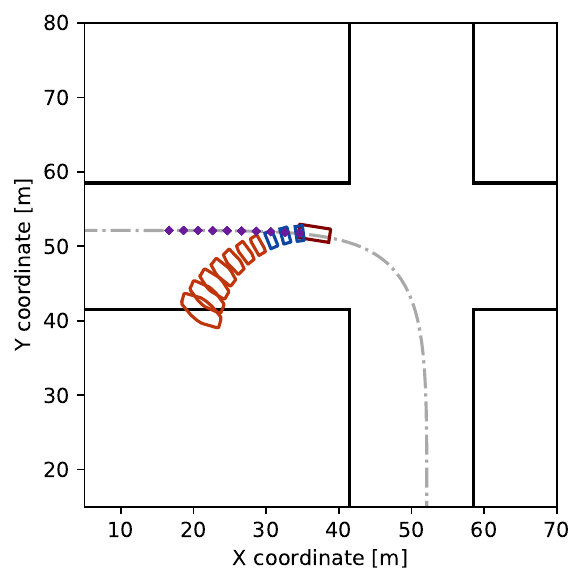}
        \caption{k = 30}
        \label{fig:scenario_30}
    \end{subfigure}
    \hfill
    \begin{subfigure}[t]{0.31\columnwidth}
        \centering
        \includegraphics[width=\linewidth]{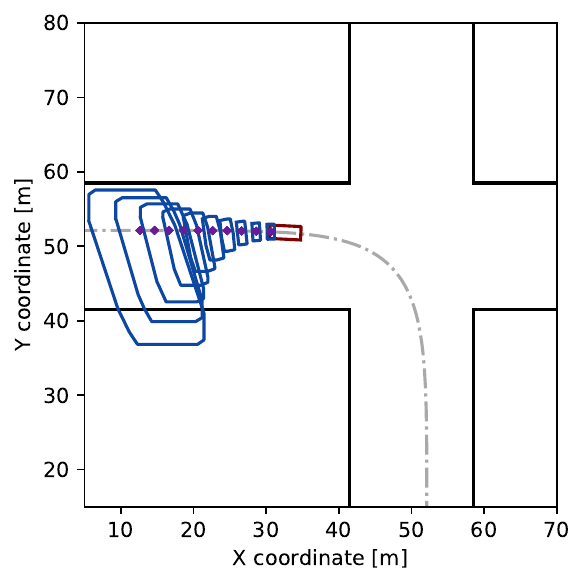}
        \caption{k = 32}
        \label{fig:scenario_32}
    \end{subfigure}
    \hfill
    \begin{subfigure}[t]{0.31\columnwidth}
        \centering
        \includegraphics[width=\linewidth]{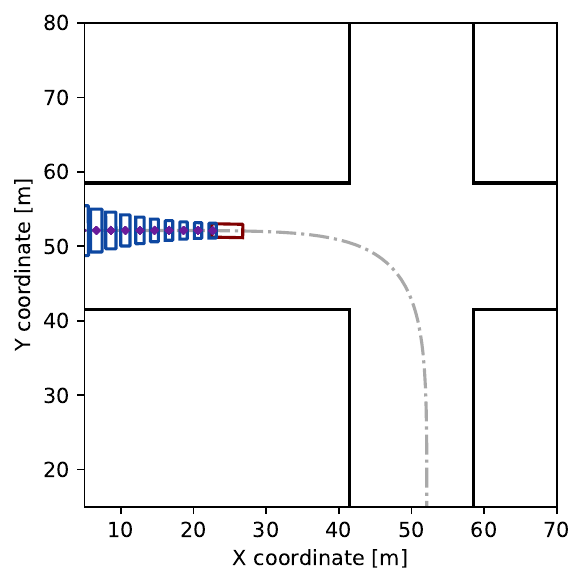}
        \caption{k = 36}
        \label{fig:scenario_36}
    \end{subfigure}
    \caption{\textbf{Proposed method.} Dark red: current SV location; purple: subsequent locations used as ground truth; blue: successfully predicted occupancy sets; orange: missed.}
    \label{fig:scenario_comparison}
\end{figure}

This subsection examines adaptation on the most critical case shown in Figure~\ref{fig:scenario_comparison}, where two consecutive maneuver changes occur, representing the period of lowest prediction accuracy.
 During a left turn (Figure~\ref{fig:scenario_12}), the SV approaches a maneuver change, and gradually moves toward the edge of the predicted occupancy set leading to mismatches in medium- and long term predictions (Figure~\ref{fig:scenario_20}). Still, the first predictions remain accurate, and the number of steps depends on the chosen performance-conservatism trade-off. Once the maneuver change is observed, the occupancy sets expand to encompass both the previous and current behaviors (Figure~\ref{fig:scenario_21}). After new behavior is confirmed, the sets progressively shrink, refining their shape and providing a tightener estimation of the behavior. The same expansion-contraction pattern appears again when the vehicle resumes straight motion, as shown in Figures~\ref{fig:scenario_30}-~\ref{fig:scenario_36}, illustrating how the method readapts once the behavior stabilizes.

\begin{figure}[htb!]
    \centering
    \begin{subfigure}[t]{0.31\columnwidth}
        \centering
        \includegraphics[width=\linewidth]{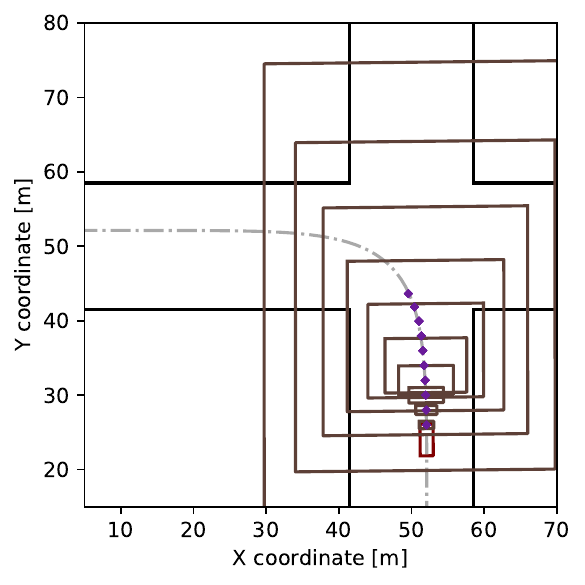}
        \caption{k = 12}
        \label{fig:wc_12}
    \end{subfigure}
    \hfill
    \begin{subfigure}[t]{0.31\columnwidth}
        \centering
        \includegraphics[width=\linewidth]{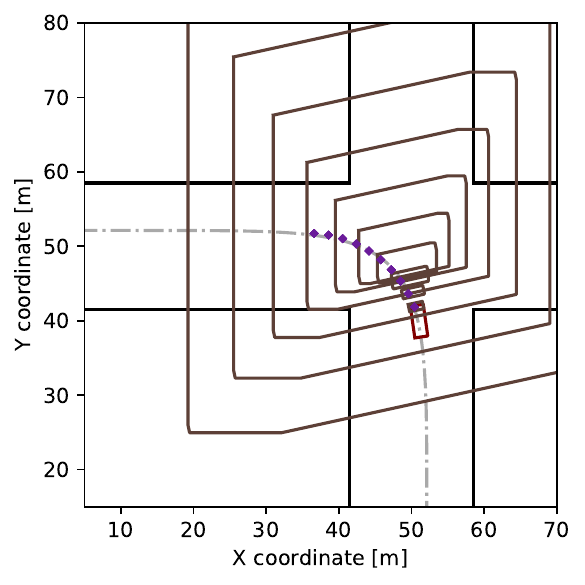}
        \caption{k = 20}
        \label{fig:wc_20}
    \end{subfigure}
    \hfill
        \begin{subfigure}[t]{0.31\columnwidth}
        \centering
        \includegraphics[width=\linewidth]{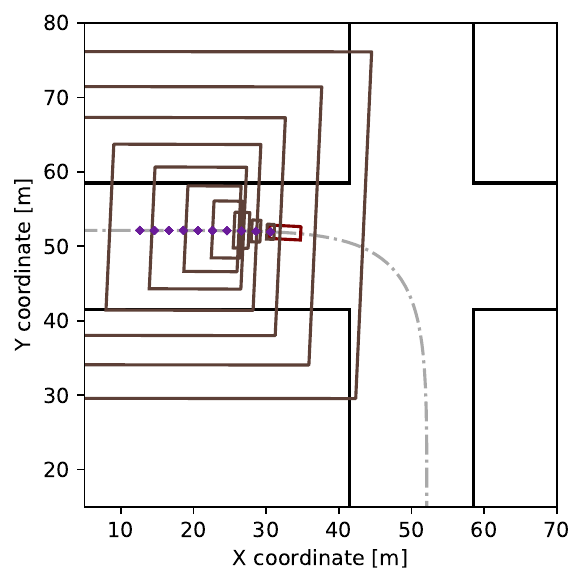}
        \caption{k = 32}
        \label{fig:wc_32}
    \end{subfigure}
    \caption{\textbf{Non-adaptive baseline.} Dark red, current SV location; purple: subsequent locations used as ground truth; brown: predicted occupancy sets.}
    \label{fig:worst_case_comparison}
\end{figure}

Finally, to compare the proposed method with a non-adaptive robust method, a comparative simulation was conducted using a fixed Control-Input set covering all possible behavior. This set was built as a bounding box enclosing all samples from Figure~\ref{fig:path_followed}. The results, shown in Figure~\ref{fig:worst_case_comparison} for the same time steps as Figures~\ref{fig:scenario_12}, \ref{fig:scenario_20}, and \ref{fig:scenario_32}, reveal that worst-case predictions are much more conservative, covering larger areas than the proposed method. This excessive over-approximation severely limits the available space, making it unsuitable for safe motion planning of the EV. 
These results clearly highlight the advantages of the proposed approach over non-adaptive strategies, especially in urban scenarios where maintaining compact occupancy sets is essential. Even when the actual maneuver diverges from expectations, the short-term prediction remains valid. More conservative sets are temporarily generated to preserve safety, but they remain significantly more compact than those from a worst-case strategy. Once the new behavior is confirmed, the method quickly adapts and refines the predictions, reducing the size of the sets accordingly. This reinforces the suitability of the method for realistic urban scenarios, where adaptiveness and compact predictions are crucial for efficient motion planning.

\section{CONCLUSIONS AND FUTURE WORK} \label{sec:conclusions}
This work presents an efficient and accurate methodology to estimate occupancy sets solely from observations. It automatically adapts the level of conservatism to the dispersion of observations dispersion, and, thanks to its zonotope-based formulation, achieves more compact sets than existing methods, such as worst-case approaches or those based on strong priors. This makes it suitable for complex and dynamic environments, such as mixed traffic urban scenarios.
At this point, different lines of future work are proposed. One is to explore alternative ways to compute the zonotope generators based on observation dispersion, aiming to reduce conservatism. Another is to integrate the estimator into an autonomous vehicle motion planner and assess its ability to generate efficient safe trajectories in comparison with existing interaction-aware planners.
\bibliographystyle{ieeetr}
\bibliography{references}
\end{document}